%% file: ms.tex
\documentclass[12pt,a4paper]{article}

\pdfoutput=1

\usepackage[margin=1in]{geometry}
\usepackage{microtype}
\usepackage{graphicx}
\graphicspath{{./figures/}}
\usepackage{subfigure}
\usepackage{booktabs} % for professional tables
\usepackage{latexsym}
\usepackage{algorithmic}
\usepackage{amssymb}
\usepackage{algorithm}
\usepackage{amssymb}
\usepackage{amsfonts}
\usepackage{float}
\usepackage{bm}
\usepackage{amsmath}
\usepackage{xcolor}
\usepackage[bottom]{footmisc}
\usepackage{authblk}
\usepackage{caption}
\DeclareMathOperator{\Tr}{Tr}

\usepackage{hyperref}

\title{Inverse Estimation of Elastic Modulus Using Physics-Informed Generative Adversarial Networks }

\author[1]{James E. Warner\thanks{Corresponding Author (\texttt{james.e.warner@nasa.gov})}}
\author[2]{Julian Cuevas}
\author[1]{Geoffrey F. Bomarito}
\author[1]{Patrick E. Leser }
\author[1]{William P. Leser}
\affil[1]{\textit{NASA Langley Research Center, Hampton, VA}}
\affil[2]{\textit{University of Puerto Rico (Mayag\"{u}ez)}, Mayag\"{u}ez, PR}

\date{}

\begin{document}
\maketitle

\abstract{While standard generative adversarial networks (GANs) rely solely on training data to learn unknown probability distributions, physics-informed GANs (PI-GANs) encode physical laws in the form of stochastic partial differential equations (PDEs) using auto differentiation.  By relating observed data to unobserved quantities of interest through PDEs, PI-GANs allow for the estimation of underlying probability distributions without their direct measurement (i.e. inverse problems). The scalable nature of GANs allows high-dimensional, spatially-dependent probability distributions (i.e., random fields) to be inferred, while incorporating prior information through PDEs allows the training datasets to be relatively small. 

In this work, PI-GANs are demonstrated for the application of elastic modulus estimation in mechanical testing. Given measured deformation data, the underlying probability distribution of spatially-varying elastic modulus (stiffness) is learned. Two feed-forward deep neural network generators are used to model the deformation and material stiffness across a two dimensional domain. Wasserstein GANs with gradient penalty are employed for enhanced stability. In the absence of explicit training data, it is demonstrated that the PI-GAN learns to generate realistic, physically-admissible realizations of material stiffness by incorporating the PDE that relates it to the measured deformation. It is shown that the statistics (mean, standard deviation, point-wise distributions, correlation length) of these generated stiffness samples have good agreement with the true distribution.}

\section{Introduction}
 \input{introduction}

\section{PI-GANs for Inverse Problems}
\label{sec:formulation}
\input{formulation}

\section{Application}
\label{sec:application}
\input{application}

\section{Numerical Example}
\label{sec:example}
\input{example}

\section{Conclusion}
\label{sec:conclusion}
\input{conclusion}

\section{Appendix}
\input{appendix}

\section*{Acknowledgement}
This work was funded by the Center Internal Research and Development (IRAD) and High Performance Computing Incubator (HPCI) programs at NASA Langley Research Center.

\bibliographystyle{plain}
\bibliography{pigans_preprint}

\end{document}

%% file: introduction.tex
The field of scientific machine learning has recently formed around the idea of infusion of current scientific knowledge into machine learning contexts \cite{baker_2019}.  This most commonly comes in the form of leveraging domain-specific knowledge in machine learning work flows \cite{roscher2019explainable}.  For example, in physics-informed (PI) machine learning, known physical constraints are used as regularization in deep learning \cite{raissi_2019}. The physical constraints are encoded as partial differential equations (PDEs) using auto differentiation. Especially in the common case where data is limited but prior knowledge of the physical system is present, the regularization can result in more robust training and more accurate models \cite{raissi_2019,tartakovsky2018learning,ling2016reynolds}.  Additionally, the resulting models can be easier to trust, based on their adherence to current scientific knowledge.

%The seminal work in generative adversarial networks (GANs) \cite{goodfellow2014} has had more than 15,000 citations in the six years since its publication. 
%Meanwhile, the use of generative adversarial networks (GANs) has seen a huge surge in recent years, with the seminal work  \cite{goodfellow2014}  having received more than 15,000 citations in the six years since its publication.
%Meanwhile,  generative adversarial networks (GANs) have seen tremendous success since their inception, with the seminal work  \cite{goodfellow2014}  having received more than 15,000 citations in the six years following its publication. 

Generative adversarial networks (GANs) have seen tremendous success since their introduction in 2014, with the seminal work  \cite{goodfellow2014}  having received more than 15,000 citations in the six years following its publication. Given the documented effectiveness of GANs for learning probability distributions from data \cite{ledig2016photorealistic,zhang2017adversarial,yang2017midinet}, recent efforts have utlized GANs for uncertainty quantification in the context of scientific machine learning \cite{yang2018physicsinformed,yang2019highly,yang_2019}.  Particularly, the ability of GANs to scale to high stochastic dimensions makes them a promising alternative to common methods such as polynomial chaos expansion \cite{ghanem_1990, xiu_2002, xiu_2010, oladyshkin_2012}, stochastic collocation \cite{babuvska_2007, nobile_2008}, and stochastic reduced order models \cite{grigoriu_2009, warner_2013, warner_2015}. Early adoption of GANs was hindered by training instability, but recent contributions such as the Wasserstein GAN (WGAN) \cite{arjovsky_2017} and the WGAN with gradient penalty (WGAN-GP) \cite{gulrajani_2017} have significantly improved convergence properties.

Earlier efforts to combine PI machine learning and GANs (PI-GANs) \cite{yang2018physicsinformed,yang_2019} have been partially motivated by their ability to perform inverse problems, i.e., inference of an unknown probability distribution based on its relation to another observed quantity. The formulations provided in these works generally define the relationship between the unknown and observed quantities as a vector-valued, stochastic PDE, and allow either or both quantities to be spatially or temporally varying. Importantly, however, the applications considered in these works either involved (i) a small, non-zero (e.g., as low as one) number of observations of the unknown quantity \cite{yang2018physicsinformed} or (ii) a direct functional dependence of the unknown quantity on the observed quantity (i.e., the unknown quantity is a function of the solution to the PDE) \cite{yang_2019}.

For many computational science and engineering inverse problems, the unknown probability distribution is specifically \emph{unobservable} and may not be functionally dependent on the solution to the PDE. Solving this class of inverse problem was the focus of this work. Here, a PI-GAN was formulated which uses WGAN-GP and explicit penalization of a vector-valued PDE and arbitrarily many boundary conditions. Without direct observations of the quantity of interest, inference is strictly dependent on the PDE residual evaluated at collocation points in the problem domain. With this in mind, the proposed formulation follows previous work \cite{raissi_2019, yang_2019} and avoids hard limits (e.g., computer memory) on the number of collocation points by including a separate PDE loss term. The PI-GAN was applied to a solid mechanics problem in which an unobservable, spatially varying material property was inferred given observations of a two-dimensional material response. To the best of the authors{'} knowledge, this is the first example of PI-GANs in the field of solid mechanics.

The remainder of this paper is organized as follows. Section \ref{sec:formulation} formally defines the inverse problem and describes the PI-GAN formulation used herein.  The problem of inverse material identification and its relevant physics are introduced in Section \ref{sec:application}. The material identification problem is then performed in Section \ref{sec:example}, wherein results on the accuracy and robustness of the method are discussed.  Finally, Section \ref{sec:conclusion} outlines contributions and conclusions.

%% file: formulation.tex
The PI-GAN framework integrates the deep learning concepts of GANs \cite{goodfellow2014} and physics-informed neural networks \cite{raissi_2019} to learn probability distributions from data while adhering to relevant physical laws. This section provides a brief overview of the formulation in the context of solving PDE-constrained inverse problems.

\subsection{Inverse Problem Definition}\label{section:inverse_problem}

Consider the following general form of a time-independent stochastic PDE
\begin{align}
	 \mathcal{N}_x [\mathbf{u}(\mathbf{x}, \omega); E(\mathbf{x}, \omega)] &=  \mathbf{f}(\mathbf{x}), \; \mathbf{x} \in \mathcal{D}, \; \omega \in \Omega \label{general_pde} \\
	 \mathcal{B}_{x}^{(k)} [\mathbf{u}(\mathbf{x}, \omega); E(\mathbf{x}, \omega)]  &= \mathbf{b}^{(k)}(\mathbf{x}), \; \mathbf{x} \in \Gamma^{(k)}, \label{general_bc}
\end{align}
where $\mathcal{N}_x$ is an arbitrary differential operator, $\mathcal{D} \subset \mathbb{R}^d$ is the physical domain, $\Omega$ is the probability space, and $\{\mathcal{B}_x^{(k)} \}_{k=1}^{N_k}$ are the $N_k$ boundary condition operators applied to the respective portions of the boundary, $\Gamma^{(k)}$. With a proper specification of the coefficient $E(\mathbf{x}, \omega)$, forcing function $\mathbf{f}(\mathbf{x})$, and boundary conditions $\{\mathbf{b}^{(k)}(\mathbf{x})\}_{k=1}^{N_k}$, the \textit{forward problem} can be solved to approximate the solution, $\mathbf{u}(\mathbf{x}, \omega)$. Here, $\mathbf{u}: \mathbb{R}^d \times \Omega \to \mathbb{R}^d$ is a vector-valued random field, i.e., a function of both the spatial coordinate, $\mathbf{x}$, and random event, $\omega$. Note that it has been assumed that uncertainty in $\mathbf{u}$ is solely due to randomness in $E$ (the forcing function and boundary condition are deterministic) for simplicity, but this need not be the case. 

This work focuses on the corresponding \textit{inverse problem} associated with Equations \eqref{general_pde} and \eqref{general_bc}. That is, to determine $E(\mathbf{x}, \omega)$ using incomplete observations of the solution $\mathbf{u}(\mathbf{x}, \omega)$, again with known functions, $\mathbf{f}(\mathbf{x})$ and $\mathbf{b}^{(k)}(\mathbf{x})$\footnote{When the solution to the inverse problem is the coefficient of a PDE, it is often referred to as parameter estimation.}. The observations of $\mathbf{u}$ are assumed to have been collected from $N_u$ independent measurements at $N_{\hat{\mathbf{x}}_u}$ sensor locations, $\{ \hat{\mathbf{x}}^{(i)}_u \}_{i=1}^{N_{\hat{\mathbf{x}}_u}}$, resulting in the dataset, 
%\begin{equation}
%	\mathcal{U} = \{ \hat{u}^{(i,j)} \}_{i=1}^{N_{\hat{\mathbf{x}}_u}}, \; j=1,...,N_u  \label{u_snapshots}
%\end{equation}	
\begin{equation}
	\mathcal{U} = \{ \hat{u}^{(i,j)} \}, \; i = 1, ..., N_{\hat{\mathbf{x}}_u},  \; j=1,...,N_u,  \label{u_snapshots}
\end{equation}	
where $ \hat{u}^{(i,j)} \equiv  \hat{u} (  \hat{\mathbf{x}}^{(i)}_u, \omega^{(j)} )$. An ideal solution to this inverse problem is one that provides a full spatial and probabilistic description of $E$ capable of producing realistic random field realizations that are physically admissible according to Equations \eqref{general_pde} and \eqref{general_bc} (as opposed to only low order statistics of $E$). 

\subsection{Generative Adversarial Networks (GANs)}\label{section:gans}

Standard generative adversarial networks (GANs) solve the problem of learning a probability distribution given sample data. For example, using a training dataset like the one in Equation \eqref{u_snapshots}, GANs can learn to produce samples that closely mimic those from the true underlying $\mathbf{u}$ distribution, $\mathbb{P}_\mathbf{u} \in \mathbb{R}^d$. Training is done through a game between two competing networks: a generator network that learns to map random noise to realistic samples of $\mathbf{u}$ and a discriminator network that learns to distinguish between generated (fake) samples and true samples from $\mathcal{U}$.

Let the generator, $G_{\theta}( \cdot )$, and discriminator, $D_{\phi}( \cdot )$, be two feed-forward deep neural networks parameterized by  $\theta$ and $\phi$, respectively. Here, the generator accepts the random variable, $\mathbf{z}  \in \mathbb{R}^m$ (with assumed distribution $\mathbb{P}_z$) as an input and outputs a sample $G_{\theta}( \mathbf{z} )  \in \mathbb{R}^d$ (with learned distribution $\mathbb{P}_g$). The goal of the generator is to approximate $\mathbb{P}_\mathbf{u}$ with $\mathbb{P}_g$. The discriminator, on the other hand, takes a sample $\mathbf{v}  \in \mathbb{R}^d$ and learns to classify it as being fake (from $\mathbb{P}_g$) or real (from $\mathbb{P}_\mathbf{u}$).

The GAN training procedure can be formally stated as:
\begin{equation}
	\underset{G}{\text{min}} \; \underset{D}{\text{max}}  \;  \mathbb{E}_{\mathbf{v} \sim \mathbb{P}_{\mathbf{u}}} [\log(D_{\phi}(\mathbf{v})) ] +   \mathbb{E}_{\mathbf{z} \sim \mathbb{P}_z} \left[\log(1 - D_{\phi}(G_{\theta}( \mathbf{z} ))) \right], \label{gan_problem}
\end{equation}
where the loss functions for the generator and discriminator are
%\begin{align}
%	\mathcal{L}_G &= \mathbb{E}_{\mathbf{z} \sim \mathbb{P}_z} \left[\log(1 - D_{\phi}(G_{\theta}( \mathbf{z} ))) \right] \label{gan_gen_loss} \\
%	\mathcal{L}_D &= -\mathbb{E}_{\mathbf{v} \sim \mathbb{P}_{\mathbf{u}}} [\log(D_{\phi}(\mathbf{v}) ] - \mathbb{E}_{\mathbf{z} \sim \mathbb{P}_z} \left[\log(1 - D_{\phi}(G_{\theta}( \mathbf{z} ))) \right].  \label{gan_disc_loss}
%\end{align}
\begin{equation}
	\mathcal{L}_G = \mathbb{E}_{\mathbf{z} \sim \mathbb{P}_z} \left[\log(1 - D_{\phi}(G_{\theta}( \mathbf{z} ))) \right] \label{gan_gen_loss}
\end{equation}
and
\begin{equation}
	\mathcal{L}_D = -\mathbb{E}_{\mathbf{v} \sim \mathbb{P}_{\mathbf{u}}} [\log(D_{\phi}(\mathbf{v}) ] -  \mathbb{E}_{\mathbf{z} \sim \mathbb{P}_z} \left[\log(1 - D_{\phi}(G_{\theta}( \mathbf{z} ))) \right],  \label{gan_disc_loss}
\end{equation}
respectively. It can be shown that if the discriminator is optimal, then minimizing Equation \eqref{gan_gen_loss} is equivalent to minimizing the Jensen-Shannon (JS) divergence between $\mathbb{P}_{\mathbf{u}}$ and $\mathbb{P}_{g}$. However, training GANs through the solution of Equation \eqref{gan_problem} is a delicate process and can be plagued with vanishing gradients and instabilities. 

Following the developments in \cite{yang2018physicsinformed},  Wasserstein GANs (WGANs) with gradient penalty  \cite{gulrajani_2017} are adopted in this work for improved training stability and approximation of $\mathbb{P}_{\mathbf{u}}$. The original GAN generator and discriminator loss functions are modified in this case as follows:
\begin{equation}
	\mathcal{L}^w_G = -\mathbb{E}_{\mathbf{z} \sim \mathbb{P}_z} \left[\bar{D}_{\phi}(G_{\theta}( \mathbf{z} )) \right]  \label{wgan_gen_loss} 
\end{equation}
\begin{equation}
	\mathcal{L}^w_D  = \mathbb{E}_{\mathbf{z} \sim \mathbb{P}_z} \left[\bar{D}_{\phi}(G_{\theta}( \mathbf{z} )) \right] - \mathbb{E}_{\mathbf{v} \sim \mathbb{P}_{\mathbf{u}}} [\bar{D}_{\phi}(\mathbf{v}) ]  +  \lambda  \mathbb{E}_{\hat{\mathbf{v}} \sim \mathbb{P}_{\hat{\mathbf{v}}}} \left[ ( \left\Vert \nabla_{\hat{\mathbf{v}}} \bar{D}_{\phi}(\hat{\mathbf{v}}) \right\Vert_2 - 1)^2 \right], \label{wgan_disc_loss}
\end{equation}
where the WGAN discriminator (typically called the \textit{critic} in this context), $\bar{D}_{\phi}$, is constrained to be a 1-Lipschitz function using the last term in Equation \eqref{wgan_disc_loss} to penalize discriminator gradients that deviate from unity. Here, $\lambda$ is the gradient penalty coefficient and samples from $ \mathbb{P}_{\hat{\mathbf{v}}}$ are generated by sampling uniformly along straight lines between pairs of points from $\mathbb{P}_{\mathbf{u}}$ and $\mathbb{P}_{g}$. 

Training WGANs corresponds to minimizing the Wasserstein-1 (or \textit{Earth Mover}) distance between $\mathbb{P}_{\mathbf{u}}$ and $\mathbb{P}_{g}$, rather than the JS divergence, and has been shown to make optimization of the generator easier \cite{arjovsky_2017}. Additionally, it was shown  \cite{yang2018physicsinformed} that WGANs with gradient penalty are more suitable for learning random functions that are deterministic at discrete points (on boundaries).

\subsection{Physics-Informed GANs (PI-GANs)}

While the WGAN with gradient penalty formulation described in the previous section can be used to learn $\mathbf{u}(\mathbf{x}, \omega)$ from a measurement dataset $\mathcal{U}$, the primary focus of this work is the recovery of $E(\mathbf{x}, \omega)$ using the same data. In this section, an approach to the inverse problem described in Section \ref{section:inverse_problem} is formulated using physics-informed GANs  (PI-GANs) \cite{yang2018physicsinformed} that combines measurement data from $\mathcal{U}$ and the prior knowledge of the governing stochastic PDEs in Equations \eqref{general_pde} and \eqref{general_bc}.

%WGANs with gradient penalty from previous section can be used to learn u from U training dataset. However, the focus of this work is specifically on the goal of inferring E through U where prior knowledge of the PDE Eq .. must be incorporated. In this section, an approach to the inverse problem described in Section [] is formulated using physics-informed GANs \cite{yang2018physicsinformed} that combines the information from the training dataset Equation [] and the governing stochastic PDEs Equation [ ] . 

%$\tilde{\mathbf{u}}_{\theta_u} (\mathbf{x}, \mathbf{\xi})$  $\tilde{E}_{\theta_E} (\mathbf{x}, \mathbf{\xi})$ 

With this approach, the random fields $\mathbf{u}(\mathbf{x}, \omega)$  and  $E(\mathbf{x}, \omega)$ are modeled using two independent deep neural networks
\begin{align}
	\tilde{\mathbf{u}}_{\theta_u} (\mathbf{x}, \mathbf{\xi})  &: \mathbb{R}^{d+m} \to \mathbb{R}^{d} \label{generator_u} \\
	\tilde{E}_{\theta_E} (\mathbf{x}, \mathbf{\xi}) &: \mathbb{R}^{d+m} \to \mathbb{R}^{1} \label{generator_E}
\end{align}
with parameters $\theta_u$ and $\theta_E$, respectively. Here, $\tilde{\mathbf{u}}_{\theta_u}$ and $\tilde{E}_{\theta_E}$ are generators that operate on a single spatial coordinate $\mathbf{x} \in \mathbb{R}^{d}$ and random noise input $\mathbf{\xi} \in \mathbb{R}^{m}$ to produce a realization of the random field at that point. It is important to emphasize that the generator $\tilde{E}_{\theta_E}$ represents the solution to the inverse problem defined in Section \ref{section:inverse_problem}.

In order to use measurement data on $\mathbf{u}$ to train the generator for $E$, the neural networks in Equations \eqref{generator_u} and \eqref{generator_E} must be related through the governing equations \eqref{general_pde} and \eqref{general_bc} during training. Motivated by the approach of physics-informed neural networks \cite{raissi_2019}, this is accomplished by applying the operators $\mathcal{N}_x$ and $\mathcal{B}_x^{(k)}$ to the networks $\tilde{\mathbf{u}}_{\theta_u}$ and $\tilde{E}_{\theta_E}$ to construct ``induced" neural networks for the PDE residual:
\begin{equation}
	\tilde{\mathbf{r}}_{\theta_{u,E}}(\mathbf{x}, \mathbf{\xi}) =  \mathcal{N}_x [\tilde{\mathbf{u}}_{\theta_u} (\mathbf{x}, \mathbf{\xi}); \tilde{E}_{\theta_E} (\mathbf{x}, \mathbf{\xi}) ] -  \mathbf{f}(\mathbf{x}) \label{NN_residual}
\end{equation}
and boundary condition discrepancy functions
\begin{equation}
	\tilde{\mathbf{b}}_{\theta_{u,E}}^{(k)}(\mathbf{x}, \mathbf{\xi})  =  \mathcal{B}_x^{(k)} [\tilde{\mathbf{u}}_{\theta_u} (\mathbf{x}, \mathbf{\xi}); \tilde{E}_{\theta_E} (\mathbf{x}, \mathbf{\xi})]  -  \mathbf{b}^{(k)}(\mathbf{x})  \label{NN_bcs}
\end{equation}
using auto differentiation \cite{DBLP:journals/corr/BaydinPR15}. Here, the shorthand notation $\theta_{u,E} \equiv [ \theta_{u} , \theta_{E}]$ has been used to indicate the explicit dependence of the induced networks $\tilde{\mathbf{r}}$ and $\tilde{\mathbf{b}}^{(k)}$ on the parameters of both of the original generators, $ \theta_{u}$ and  $\theta_{E}$. 

Now, the generator loss function can be augmented to encourage jointly generated samples from $\tilde{\mathbf{u}}$ and $\tilde{E}_{\theta_E}$ to satisfy the PDE and boundary conditions during training. For instance, the contribution to the loss for the PDE residual is given by
\begin{equation}
	\mathcal{L}_{PDE}(\theta_{u, E})  = \frac{1}{N_r N_{\hat{\mathbf{x}}_r}} \sum_{j=1}^{N_r} \sum_{i=1}^{N_{\hat{\mathbf{x}}_r}}  \| \hat{\mathbf{r}}_{\theta_{u,E}}^{(i,j)} \|^2, \label{pde_loss}
\end{equation}
where $\hat{\mathbf{r}}_{\theta_{u,E}}^{(i,j)} = \tilde{\mathbf{r}}_{\theta_{u,E}}(\hat{\mathbf{x}}^{(i)}_r, \mathbf{\xi}^{(j)})$ and $\{ \mathbf{x}_r^{(i)} \}_{i=1}^{N_{\hat{\mathbf{x}}_r}} \in \mathcal{D}$ are collocation points where the PDE is enforced. Similarly, the boundary condition loss is given by
\begin{equation}
	\mathcal{L}_{BC}(\theta_{u, E})  = \sum_{k=1}^{N_k} \frac{1}{N_b N_{\hat{\mathbf{x}}_{bk}}} \sum_{j=1}^{N_b} \sum_{i=1}^{N_{\hat{\mathbf{x}}_{bk}}}  \| \hat{\mathbf{b}}_{\theta_{u,E}}^{(i,j,k)} \|^2,  \label{bc_loss}
\end{equation}
where  $\hat{\mathbf{b}}_{\theta_{u,E}}^{(i,j,k)} = \tilde{\mathbf{b}}_{\theta_{u,E}}^{(k)}(\hat{\mathbf{x}}^{(i)}_{bk}, \mathbf{\xi}^{(j)})$ and $\{ \mathbf{x}_{bk}^{(i)} \}_{i=1}^{N_{\hat{\mathbf{x}}_{bk}}} \in \Gamma^{(k)}$ are the boundary condition collocation points for boundary condition $k$. Here, $N_r$ and $N_b$ are the number of randomly generated samples used to evaluate the PDE and boundary condition loss, respectively. 

The complete loss functions for PI-GANs are then as follows
\begin{align}
	&\mathcal{L}^{PI}_G(\theta_{u,E}) = \mathcal{L}^{w}_G(\theta_{u}) + \mathcal{L}_{PDE}(\theta_{u,E}) +  \mathcal{L}_{BC}(\theta_{u,E}) \label{pigan_gen_loss} \\
	&\mathcal{L}^{PI}_D(\theta_{u,E}, \phi) = \mathcal{L}^{w}_D(\theta_{u,E}, \phi). \label{pigan_disc_loss}
\end{align}
Note that the explicit dependence of the PI-GAN generator and discriminator loss functions on the parameters $\theta_E$ is what makes the solution of the inverse problem possible. In this way, $\tilde{E}_{\theta_E}$ is encouraged to learn from $\tilde{\mathbf{u}}_{\theta_u}$  through the governing equations, while $\tilde{\mathbf{u}}_{\theta_u}$ learns from the measurement data. The terms $\mathcal{L}_{PDE}$ and $\mathcal{L}_{BC}$ also act as a regularization mechanism for learning $\mathbf{u}$ by substantially restricting the space of admissible functions to only those that satisfy the governing equations.

%% file: application.tex
The problem of material identification is common to the disciplines of solid mechanics and mechanical testing. Specifically, the focus is on the estimation of spatially varying material properties -- in this case, the elastic modulus $E(\mathbf{x}, \omega)$ -- from measured deformations $\mathbf{u}(\mathbf{x}, \omega)$ under load.  The underlying probability distribution of material properties is assumed to be inherent to the material, and once identified, can aid in the quantification of uncertainty in mechanical behavior of all structures made of that material.

In this application, the PDE represented by Equation \eqref{general_pde} is:
\begin{align}
\nabla \cdot \boldsymbol{\sigma}(\mathbf{u}, E) &= \mathbf{0}, \; \mathbf{x} \in \mathcal{D}, \; \omega \in \Omega,  \label{pde_2d}
\end{align}
where $\boldsymbol{\sigma}$ is the second order tensor for stress. Assuming a 2-D plane-stress formulation, isotropic elasticity, and small strain theory \cite{reddy_2013}, the stress can be written as a function of $E$ and $\mathbf{u}$ as follows:
\begin{equation}
\boldsymbol{\sigma} = \frac{2E}{1+\nu}\left[\nabla\mathbf{u} + \nabla\mathbf{u}^T + \frac{2\nu}{1-\nu}\Tr(\nabla\mathbf{u}) \mathbf{I} \right], \label{stress_tensor}
\end{equation}
where $\nu$ is the Poisson ratio, $\Tr(\cdot)$ is the trace function, and $\mathbf{I}$ is the identity tensor.  In all cases in this work, $\nu$ is assumed to be constant ($\nu=0.3$), though the more general case of $\nu = \nu(\mathbf{x}, \omega)$ could be considered as well.

The boundary conditions are divided into two types: Dirichlet boundary conditions with known deformation $\mathbf{u}_{D}$ and Neuman boundary conditions with applied tractions $\tau$,
%\begin{align}
%\mathbf{u} &= \mathbf{u}_{D}, \; \mathbf{x} \in \Gamma_D   \label{bc_2d_1}\\
%\boldsymbol{\sigma}(\mathbf{u}, E) \cdot \mathbf{n} &= \tau, \; \mathbf{x} \in \Gamma_N \label{bc_2d_2}
%\end{align}
\begin{equation}
\mathbf{u} = \mathbf{u}_{D}, \; \mathbf{x} \in \Gamma_D   \label{bc_2d_1}
\end{equation}
and
\begin{equation}
\boldsymbol{\sigma}(\mathbf{u}, E) \cdot \mathbf{n} = \boldsymbol{\tau}, \; \mathbf{x} \in \Gamma_N, \label{bc_2d_2}
\end{equation}
where $\mathbf{n}$ is the outward normal for the surface. 

 Note that the explicit dependence of $\mathbf{u}(\mathbf{x}, \omega)$ and $E(\mathbf{x}, \omega)$ on $\mathbf{x}$ and $\omega$ in the above equations has been dropped to simplify notation.   See the Appendix for the explicit, expanded forms of the PDE and boundary conditions that are enforced through auto differentiation for PI-GAN training.

%% file: example.tex
The PI-GAN approach for inverse problems is now demonstrated for the identification of a spatially and randomly varying elastic modulus, motivated by the application described in the previous section. The framework was implemented in Tensorflow v2.0 and executed on one NVIDIA Tesla V100 GPU. All data and code associated with this work will be made available at https://github.com/NASA/pigans-material-ID. 

\subsection{Physical Problem Description}

\begin{figure} 
\centering 
\subfigure[]{\includegraphics{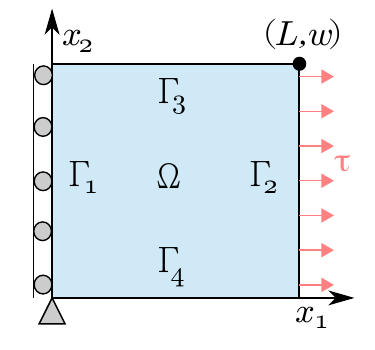} \label{domain_diagram}} 
\subfigure[]{\includegraphics{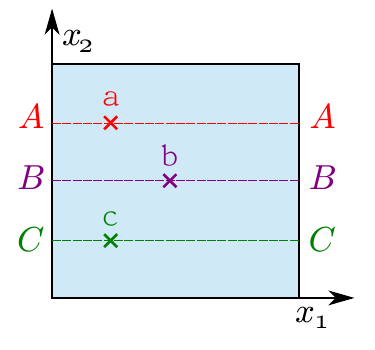} \label{plot_diagram}}
\caption{Domain for the numerical example along with (a) the boundary conditions and (b) the points and lines of interest for evaluating the accuracy of the PI-GAN.} 
\end{figure}

The problem domain can be seen in Figure \ref{domain_diagram}, assumed to be a unit square ($d = 2$) in this example ($L=w=1.0$). Dimensionless quantities are assumed throughout. Dirichlet boundary conditions (Equation \eqref{bc_2d_1}) are applied to the left boundary as follows
%\begin{align}
%	&u_1 = 0, \; 	\mathbf{x} \in \Gamma_1 \\
%	&u_2= 0,  \;	\mathbf{x}  = (0, 0),
%\end{align}
\begin{equation}
	u_1 = 0, \; 	\mathbf{x} \in \Gamma_1 \label{example_u_bc1}
\end{equation}
and
\begin{equation}
	u_2= 0,  \;	\mathbf{x}  = (0, 0), \label{example_u_bc2}
\end{equation}
where $u_1$ and $u_2$ are the components of deformation in the $x_1$ and $x_2$ directions, respectively (i.e., $\mathbf{u} = [u_1, u_2]$). Neumann boundary conditions (Equation \eqref{bc_2d_2}) are applied to the remaining boundaries, where $\boldsymbol{\tau} = [0, 0]$ on $\Gamma_3$ and $\Gamma_4$ and $\boldsymbol{\tau} = [1.5, 0]$ on $\Gamma_4$. See the Appendix for explicit forms of the implemented boundary conditions.

The true elastic modulus is modeled as a lognormal-distributed random field
\begin{align}
	E(\mathbf{x}, \omega) &= \alpha + \beta \exp( g(\mathbf{x}, \omega)) \label{lognormal_field} \\
	g(\mathbf{x}, \omega) &\sim \mathcal{G}\mathcal{P} \left(0, \exp \left(- \frac{\Vert \mathbf{x} - \mathbf{x}' \Vert ^2}{2 l^2} \right) \right)  \label{gaussian_field}
\end{align}
i.e., a zero-mean Gaussian process transformed using an exponential function and scaled using $\alpha = 1.0$ and $\beta=0.1$. The correlation length of the Gaussian process is chosen to be $l=1.0$. Samples of elastic modulus are generated by computing a truncated Karhunen-Loeve (KL) expansion \cite{10.1109/34.41390} for $g(\mathbf{x})$ and then applying the transformation in Equation \eqref{lognormal_field}. In this example, five terms are used for the KL expansion which retains $99.9\%$ of the variance of $g(\mathbf{x})$. 

The training dataset  (Equation \eqref{u_snapshots}) is generated by solving Equation \eqref{pde_2d} with the boundary conditions listed above using the finite element method \cite{hughes2000finite} as implemented in \texttt{FEniCS} Python library \cite{AlnaesBlechta2015a}. Here, a displacement solution, $\mathbf{u}$, is computed  for $N_u$ randomly sampled $E$ fields and then interpolated to $N_{\hat{\mathbf{x}}_u}$ sensor locations. In all results shown,  $N_{\hat{\mathbf{x}}_u} = 90$, and the sensor locations are chosen by creating an equidistant $10 \times 10$ grid throughout the domain and then removing the sensors along $x_1=0$. The boundary condition loss in Equation \eqref{bc_loss} is used to enforce the displacement boundary conditions (Equations \eqref{example_u_bc1} and \eqref{example_u_bc1}) on boundary $\Gamma_1$ instead.

\begin{figure} 
\centering 
\includegraphics[width=4in]{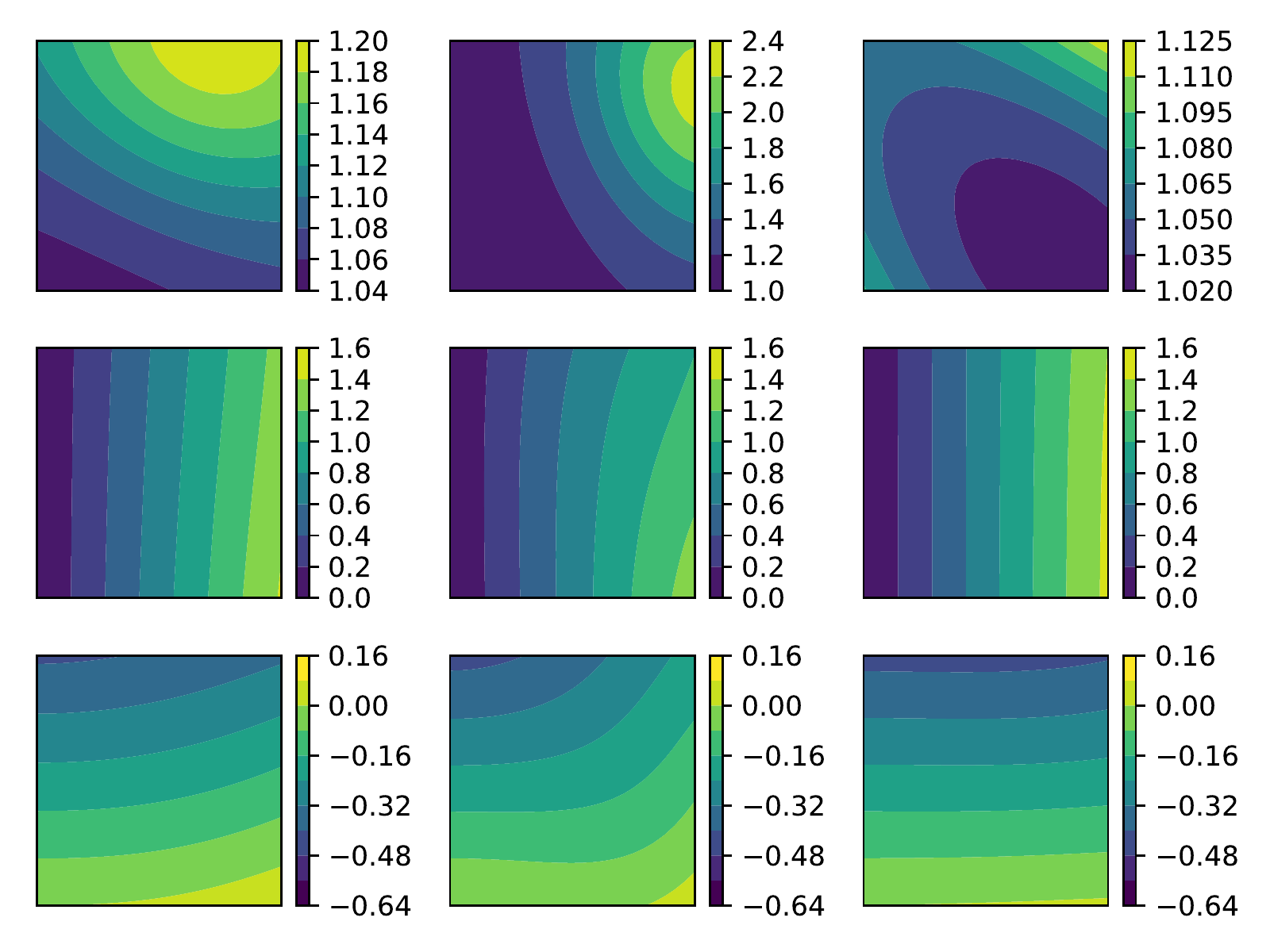}
\caption{Three randomly drawn samples of elastic modulus $E$ (row 1) and corresponding computed deformations, $u_1$ (row 2) and $u_2$ (row 3), from the test dataset.} \label{reference_samples}
\end{figure}

For illustration, Figure \ref{reference_samples} shows three representative random samples of $E$ generated using Equations \eqref{lognormal_field}, \eqref{gaussian_field}, and the KL expansion along with the corresponding solutions for $u_1$ and $u_2$ using \texttt{FEniCS}. 

\subsection{Neural Network Specification $\&$ Hyperparameters}

For all results to be shown, the generators and discriminators are implemented using feed-forward deep neural networks with four hidden layers of width 128. The hyperbolic tangent (\textit{tanh}) activation function is used for all networks, which is shown to be effective in PI-GAN approaches where higher order derivatives are necessary \cite{yang2018physicsinformed}. The random noise input $\mathbf{\xi} \in \mathbb{R}^{m}$ to the generators is distributed according to a standard multivariate Gaussian distribution with $m = 5$, so that the total input dimension to the generator is seven.

The Adam optimizer \cite{kingma2014method} is used for the minimization of the PI-GAN loss functions in Equations \eqref{pigan_gen_loss} and \eqref{pigan_disc_loss} with hyper-parameters $\beta_1 = 0$,  $\beta_2=0.9$, and learning rate of $10^{-4}$. Five generator training steps are performed for every one discriminator step  during training. A gradient penalty coefficient of $\lambda = 0.1$ is used and $10^5$ steps are taken overall. The batch size is equal to the total number of measurements in each case.

%\begin{figure} 
%\centering 
%\includegraphics[width=2.4in]{E_pdf_comparison.pdf} 
%\caption{PDF Comparison} \label{pdf_comparison}
%\end{figure}
%
%\begin{figure} 
%\centering 
%\includegraphics[width=2.4in]{E_correlation_comparison.pdf} 
%\caption{Correlation Comparison} \label{pdf_comparison}
%\end{figure}

\subsection{PI-GAN Material Identification Accuracy}\label{pigan_accurary}

First, PI-GANs were trained using $N_u = 1000$ measurements and $N_{\hat{\mathbf{x}}_r}=100$ collocation points defined on an equidistant $10 \times 10$ grid throughout the domain. One hundred generated samples were used to enforce both the PDE and boundary condition loss ($N_r = N_b = 100$) in Equations \eqref{pde_loss} and \eqref{bc_loss}, while ten equidistant points were used on each boundary ($N_{\hat{\mathbf{x}}_{bk}}$ = 10).  Accuracy in the recovery of elastic modulus was assessed by generating 1000 samples from $\tilde{E}_{\theta_E}$ after training was complete and comparing with 10000 samples from an independent test set. The average time per training step in this example was $2.5 \times 10^{-1}$ seconds, resulting in a total training time of about $7$ hours. 

\begin{figure} 
\centering 
\includegraphics[width=4in]{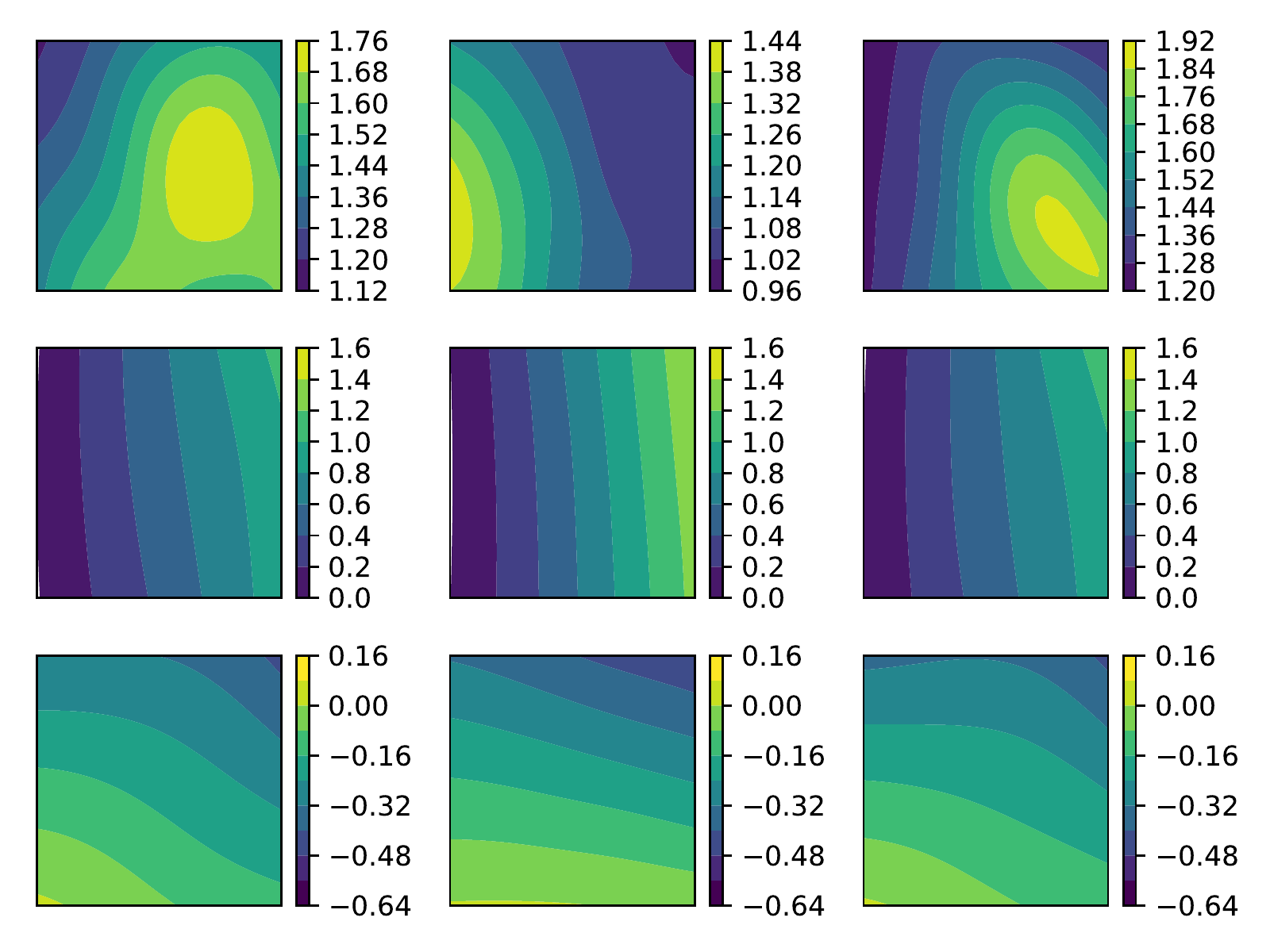}
\caption{Three randomly generated samples of  elastic modulus $E$ (row 1) and corresponding deformations, $u_1$ (row 2) and $u_2$ (row 3),  from the trained generators in Section \ref{pigan_accurary}.} \label{generated_samples}
\end{figure}

Three representative samples of $E$, $u_1$, and $u_2$ from the trained generators, $\tilde{E}_{\theta_E}$ and $\tilde{\mathbf{u}}_{\theta_u}$, are shown in Figure \ref{generated_samples}. Qualitatively speaking, the generated samples have similar overall appearance and characteristics (e.g., spatial variation and magnitude) compared to the reference samples from the test set shown in Figure \ref{reference_samples}. Note that since the sets of samples in Figures \ref{reference_samples} and  \ref{generated_samples} are drawn randomly and independently of one another, exact agreement is not to be expected. However, there will be more similarity between random realizations of $u_1$ and $u_2$ compared to $E$ due to the imposed boundary conditions.

\begin{figure} 
\centering 
\subfigure[]{\includegraphics[width=2in]{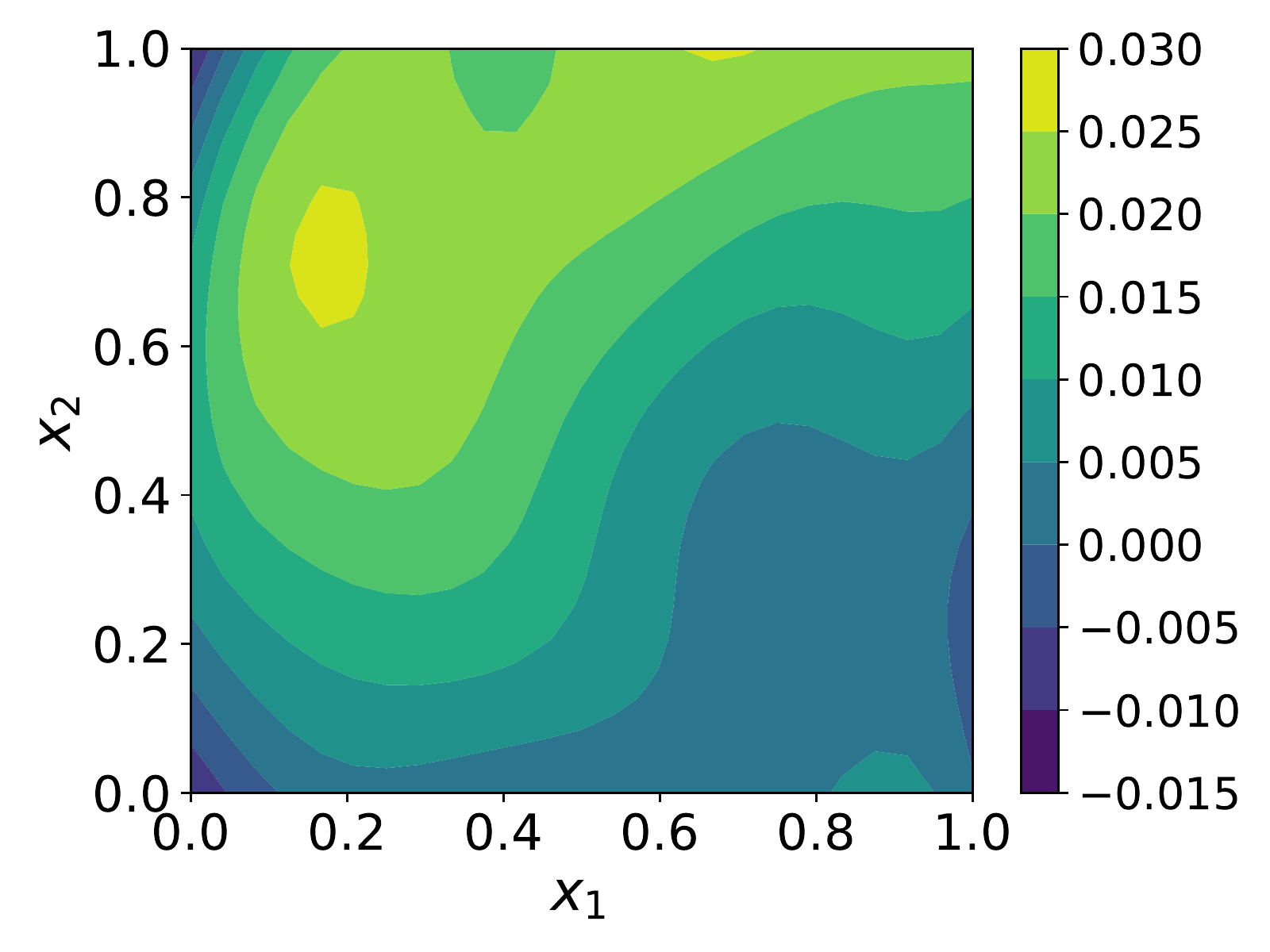}}
\subfigure[]{\includegraphics[width=2in]{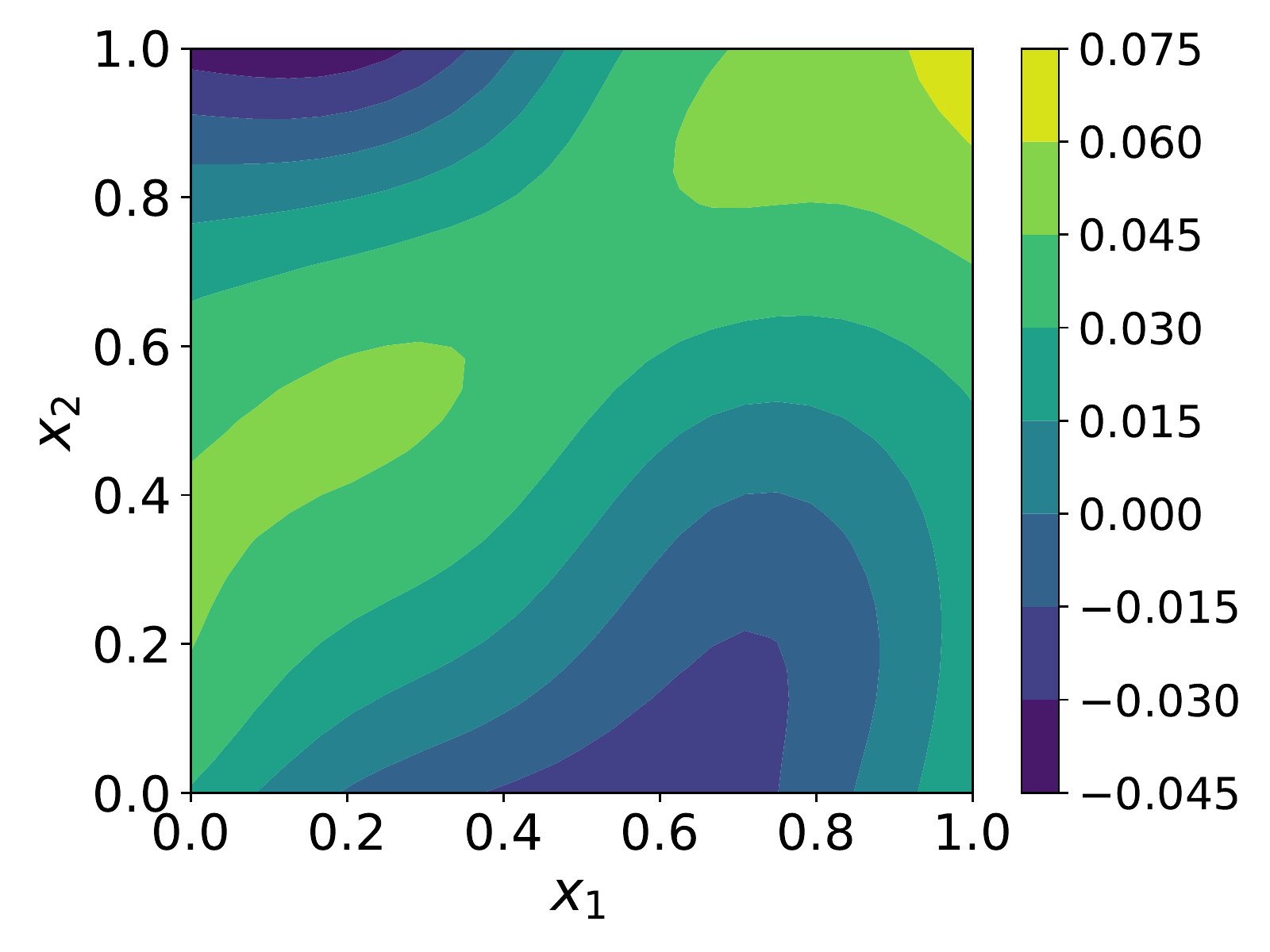}}
\caption{Point-wise error (absolute difference) between the reference and generated (a) mean and (b) standard deviation for elastic modulus.} \label{mean_std_error_contour}
\end{figure}

To assess the accuracy more quantitatively, the pointwise error in the estimated mean and standard deviation of $E$ with respect to the samples from the test set can be seen in Figure \ref{mean_std_error_contour}, . The errors for both are relatively low ($E$ has a magnitude that is $O(1)$). Better accuracy is observed for estimating the mean compared to the standard deviation, as expected.

\begin{figure} 
\centering 
\subfigure[]{\includegraphics[width=2.1in]{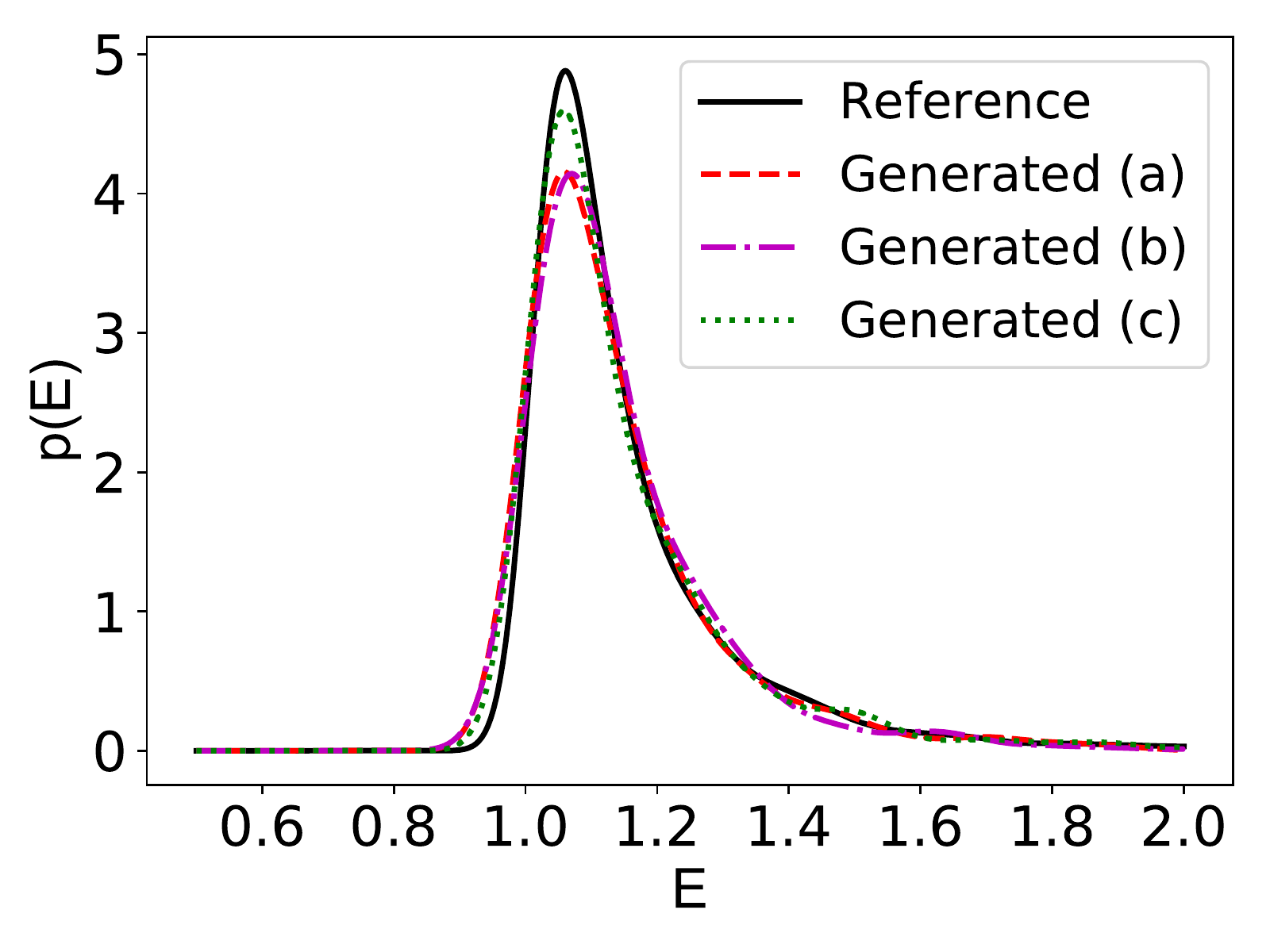} \label{pdf_comparison}} 
\subfigure[]{\includegraphics[width=2.1in]{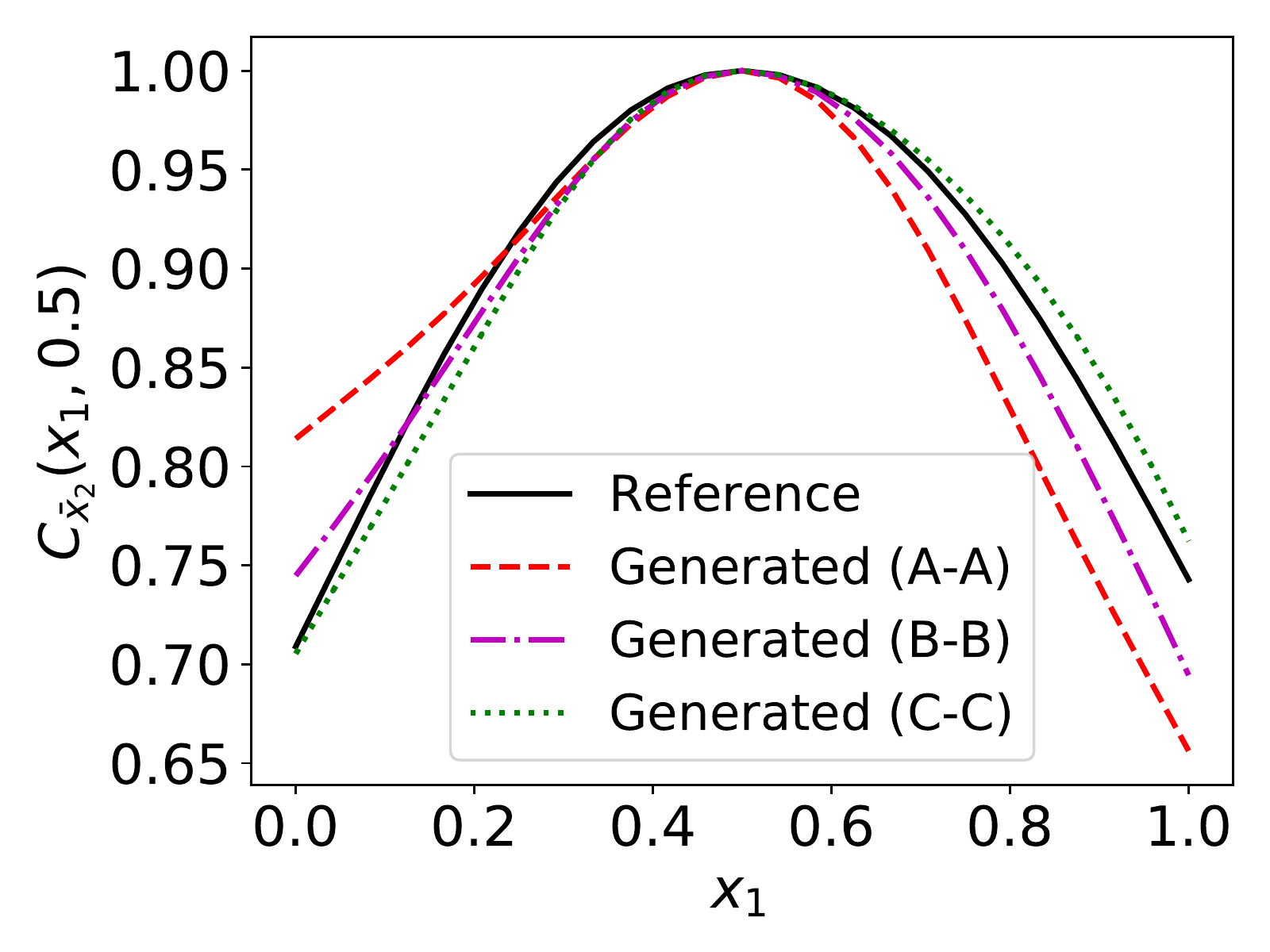} \label{correlation_comparison}}
\caption{(a) Comparison of the generated probability density function for $E$ at the points of interest in Figure \ref{plot_diagram} and (b) Comparison of the 1D correlation function (Equation \eqref{correlation1d}) across the lines of interest in Figure \ref{plot_diagram}.} 
\end{figure}

Since one primary motivation behind adopting a PI-GAN approach is to be able to capture higher order statistics beyond just mean and standard deviation, the ability of $\tilde{E}_{\theta_E}$ to recover pointwise distributions and pairwise correlations was also assessed. These comparisons were done for the points and lines of interest shown in Figure \ref{plot_diagram}. First, the estimated probability density function of $E$ using generated samples at the three reference points is shown in Figure \ref{pdf_comparison}. Good agreement is observed with the reference distribution computed from the samples in the test dataset. Note that the distribution of $E$ according to Equation \eqref{lognormal_field} is homogenous in space, so the distribution at all points should coincide\footnote{The reference distribution in Figure \ref{pdf_comparison} was computed at $\mathbf{x} = [0.5, 0.5]$.}.

Next, the accuracy of the estimated spatial correlation is shown in Figure \ref{correlation_comparison}. Here, the 1D correlation at fixed $x_2$ coordinates, 
\begin{equation}
 	C_{\bar{x}_2}(x_1, x'_1)  = \text{Correlation}\left[E(x_1, \bar{x}_2), E(x'_1 , \bar{x}_2)\right], \label{correlation1d}
\end{equation} is computed and displayed as a function of $x_1$ and compared to the reference correlation from test samples. Note that $x'_1=0.5$ and $\bar{x}_2 = 0.75, 0.5, 0.25$ for sections $A\text{-}A$, $B\text{-}B$, and $C\text{-}C$ in Figure \ref{plot_diagram}, respectively. The generated correlations decay from $C_{\bar{x}_2} = 1.0$ at $x_1 = 0.5$ towards the boundaries in good agreement with the reference solution, displaying the ability of the PI-GAN to accurately infer the spatial variation of the elastic modulus random field.

\subsection{Effect of Number of Measurements and Collocation Points}

The accuracy of the PI-GAN approach for varying numbers of measurements ($N_u$) and collocation points ($N_r$) is now illustrated. Note that all other problem specifications remain unchanged from the previous section. Three random trials were performed for each case and the average relative $L2$ error in the estimated mean and standard deviation of $E$ was computed. Here, the $L2$ norms were calculated using numerical integration on a uniform  $25 \times 25$ grid.

\begin{figure} 
\centering 
\subfigure[]{\includegraphics[width=2.25in]{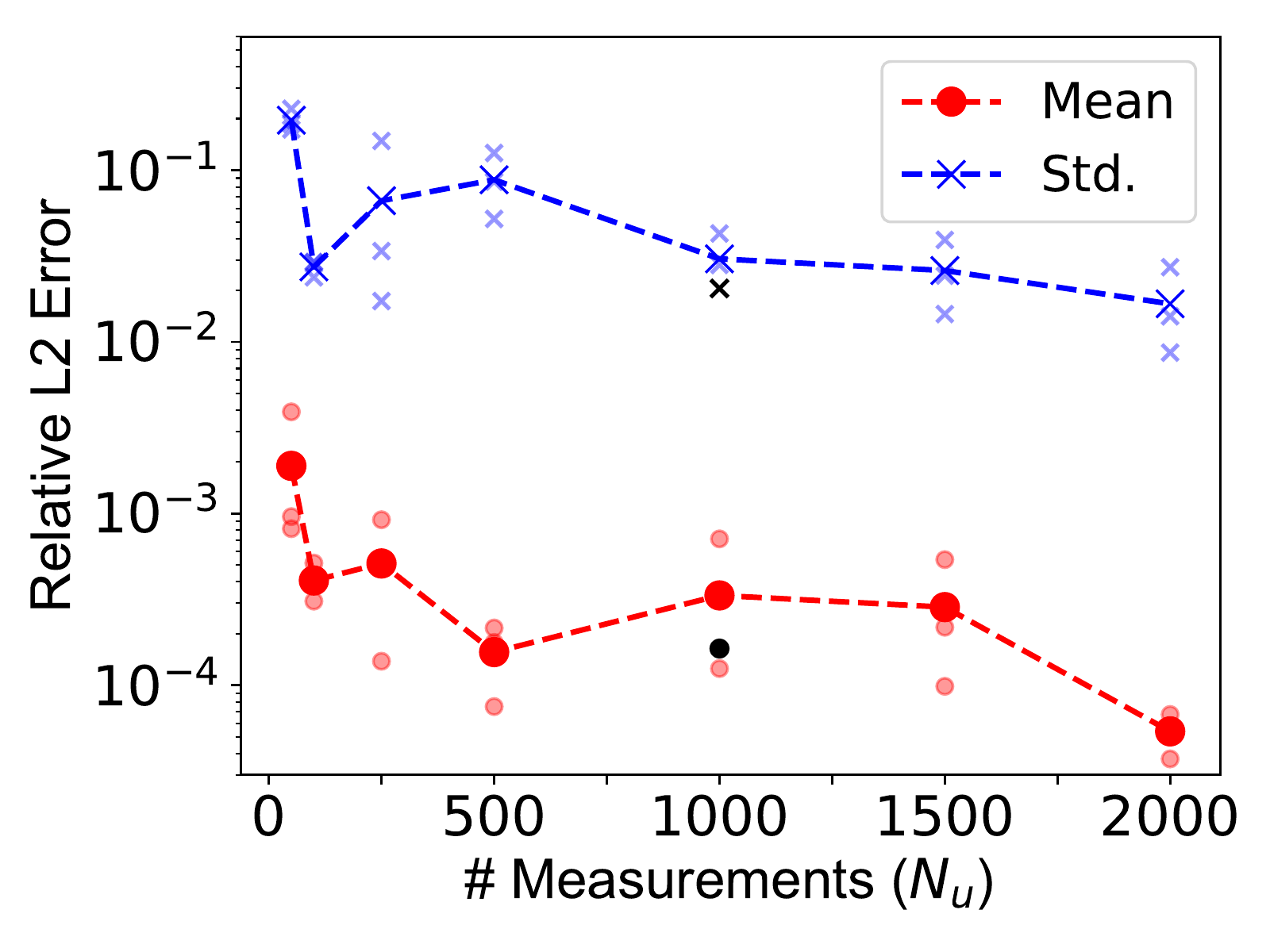} \label{error_vs_snapshot}}
\subfigure[]{\includegraphics[width=2.25in]{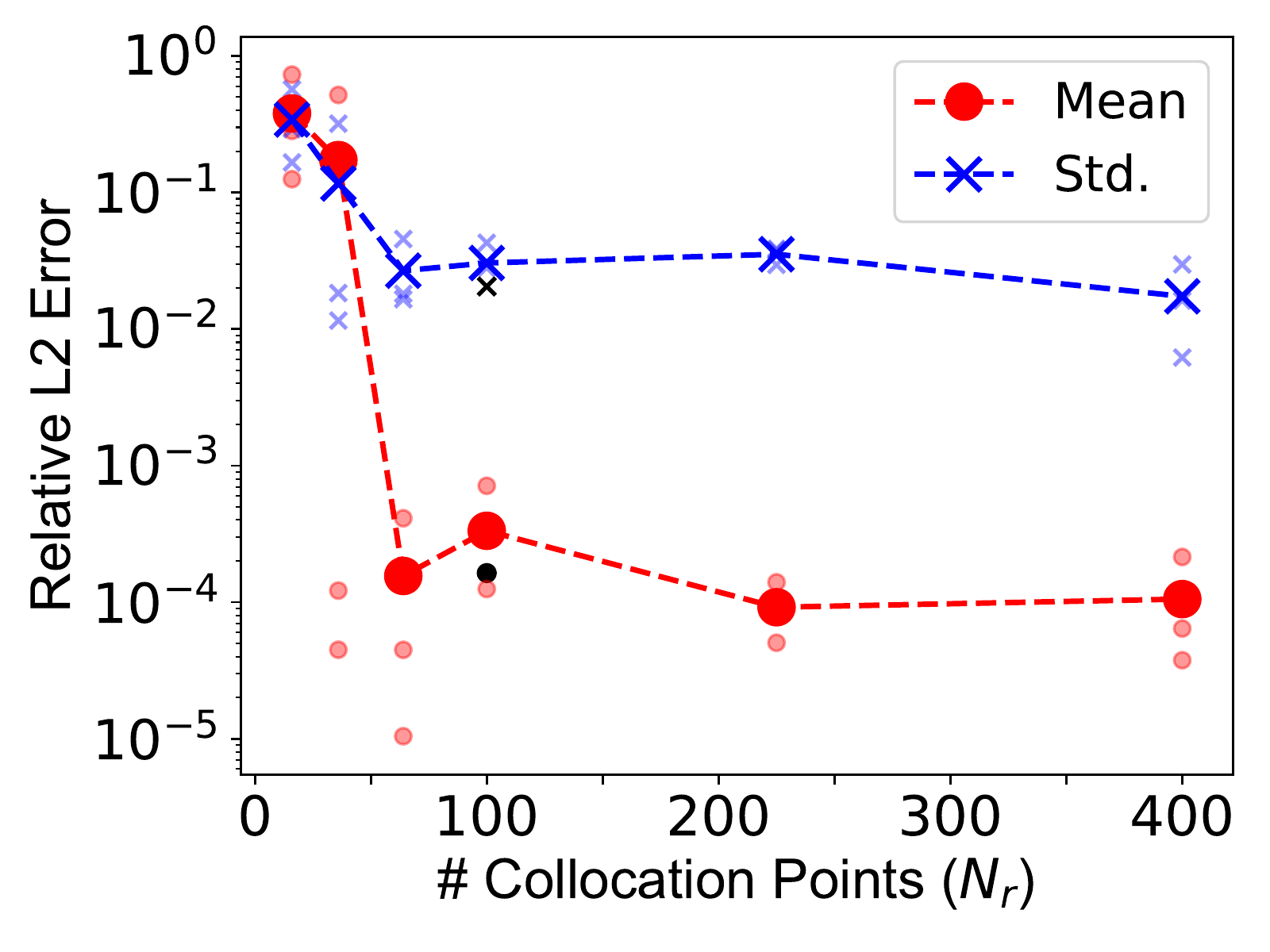} \label{error_vs_collocationpts}}
\caption{Relative L2 error in the estimated mean and standard deviation of $E$ versus (a) number of measurements and (b) number of collocation points. The curves show the average of errors from three random trials, represented individually by the smaller markers. The black symbols at $N_u=1000$ and $N_r = 100$ correspond to errors from the trial used to produce the results in Section \ref{pigan_accurary}.}
\end{figure}

First, using $N_r = 100$, the errors for varying numbers of measurements is shown in Figure \ref{error_vs_snapshot}. The errors generally decrease with increasing $N_u$ but it can be seen that relatively accurate estimates are possible with $O(100)$ measurements. Note that the black symbols at $N_u = 1000$  on the plot correspond to errors associated with the results shown in the previous section for illustration purposes.

Next, for a fixed number of measurements, $N_u = 1000$, the errors are shown versus number of collocations points in Figure \ref{error_vs_collocationpts}. Similar to the previous section, the collocation points are specified on equidistant grids across the domain: $4 \times 4$, $6 \times 6$, $8 \times 8$, $10 \times 10$, $15 \times 15$, and $20 \times 20$. As expected, the accuracy of the estimated elastic modulus field generally increases with the number of collocation points used. Here, it is important to reiterate that the primary source of information the PI-GAN has to infer $E$ is through the PDE loss in Equation \eqref{pde_loss}, and hence a sufficient number of collocation points must be used ($\gtrapprox 50$ in this example), after which diminishing returns are observed. 

Finally, it is noted that relatively significant random variation is observed in the accuracy among independent trials, as evidenced by the spread in the small markers for individual cases in Figures \ref{error_vs_snapshot} and \ref{error_vs_collocationpts}. The fact that the relative $L2$ errors are not strictly decreasing as a function of $N_u$ and $N_r$ is likely a result of this variability in training. A more thorough study on the impact of the various PI-GAN parameters (discriminator/generator architectures, number of displacement sensors, Adam optimizer hyper-parameters, WGAN gradient penalty coefficient, etc.)  on the accuracy and stability of training will be pursued in future work.

%% file: conclusion.tex
Physics-informed generative adversarial networks (PI-GANs) are an emerging tool for scientific machine learning that enable the solution of complex, stochastic partial differential equations. In this work, a PI-GAN is formulated for the purpose of solving inverse problems where an unknown probability distribution is estimated based on its relation to another, observed quantity. Specifically, this approach enables solutions to the class of inverse problem in which the unknown quantity is strictly unobservable and is not a function of the PDE solution. As a result, the only information used to train the feed-forward neural network representing the unknown probability distribution is indirectly obtained through the governing PDE and boundary conditions. To address this, separate PDE and boundary condition loss terms are included in the classic generator-discriminator GAN formulation. Additionally, recent improvements regarding training stability are incorporated by using the Wasserstein GAN with gradient penalty (WGAN-GP). The formulation allows for vector-valued PDEs with unknown and spatially varying quantities of interest and for arbitrary boundary conditions.

The PI-GAN was demonstrated on a solid mechanics example in which the unknown, spatially varying stiffness distribution was estimated given two-dimensional observations of material deformation in response to an applied load. In the context of this numerical example, it was shown that statistics (mean, standard deviation, point-wise distributions and correlation length) of the generated stiffness samples were in good agreement with those of the true distribution. Additionally, the effect on accuracy of varying the number of observations and collocation points (i.e., the number of points at which the PDE constraint was enforced in the problem domain during training) was studied. It is found that $O(100)$ measurements were enough to produce accurate estimates given $100$ collocation points. Fixing the number of measurements at $1000$, accurate estimates were achieved with $O(10)$ collocation points ($\gtrapprox 50$ in the specific example presented herein). To the best of the authors' knowledge, this is both the first application of a PI-GAN in the field of solid mechanics and to this specific class of inverse problem (i.e., no direct observations of the quantity of interest and no dependence of that quantity on the PDE solution).

%% file: appendix.tex
\textit{PDE Residual Expressions} 

A more explicit form of the PDE residual implemented in this work can be obtained by substituting Equation \eqref{stress_tensor} into Equation \eqref{pde_2d} and expressing the result in index notation, i.e.,
\begin{multline}
	\frac{(1-\nu)}{2} \left[ E_{,2}  (u_{1,2} + u_{2,1} ) + E  (u_{1,22} + u_{2,12} ) \right]  +  \\ E_{,1} \left(u_{1,1} + \nu u_{2,2} \right) + E \left(u_{1,11} + \nu u_{2,21} \right) = 0  \label{pde_constraint1_explicit} 
\end{multline} 
and
\begin{multline}
	\frac{(1-\nu)}{2}  \left[ E_{,1} (u_{1,2} + u_{2,1} ) + E (u_{1,21} + u_{2,11} ) \right] +  \\ E_{,2} \left( \nu u_{1,1}  +  u_{2,2} \right)  + E \left( \nu u_{1,12}  +  u_{2,22} \right)  = 0,  \label{pde_constraint2_explicit} 
\end{multline} 
where the notation $f_{i,j} \equiv \frac{\partial f_i}{\partial x_j}$ has been adopted.

\textit{Boundary Condition Expressions} 

A more explicit form of the implemented Neumann boundary conditions from the numerical example can be derived by using $\mathbf{n} = [1, 0]$ on $\Gamma_2$, $\mathbf{n} = [0, 1]$ on $\Gamma_3$, and $\mathbf{n} = [0, -1]$ on $\Gamma_4$ with Equation \eqref{bc_2d_2} and expressing the result in index notation, i.e.,
\begin{align}
	 \frac{E}{1-\nu^2} \left(u_{1,1} + \nu u_{2,2} \right) &= 1.5 \;\; \mathbf{x} \in \Gamma_2, \\
	 \frac{E}{2(1+\nu)} \left(u_{1,2} + u_{2,1} \right) &= 0 \;\; \mathbf{x} \in \Gamma_2, \\
	 \frac{E}{1-\nu^2} \left(\nu u_{1,1} + u_{2,2} \right) &= 0 \;\; \mathbf{x} \in \Gamma_3,  \Gamma_4, \; \text{and}\ \\
	 \frac{E}{2(1+\nu)} \left(u_{1,2} + u_{2,1} \right) &= 0 \;\; \mathbf{x} \in \Gamma_3, \Gamma_4.
\end{align}

For more implementation details, the interested reader can find the code and data needed to reproduce the results of this report at https://github.com/NASA/pigans-material-ID.

%\begin{equation}
%	\left[\frac{\partial}{\partial x} \; \;  \frac{\partial}{\partial y}  \right] \cdot 
% 	\begin{bmatrix} 
%		\sigma_{xx} & \sigma_{xy} \\
%		\sigma_{yx} & \sigma_{yy} 
%	\end{bmatrix}
%	= \left[ 0  \;\;  0 \right] ,  \label{pde_2d_matrixform}
%\end{equation}
%
% \begin{equation}
%	 \begin{bmatrix} 
%		\sigma_{xx}  \\
%		\sigma_{yy}  \\
%		\sigma_{xy} 
%	\end{bmatrix}
%	=  \frac{E(x,y)}{1-\nu^2}
%	 \begin{bmatrix} 
%		1 & \nu & 0  \\
%		\nu & 1 & 0  \\
%		0 & 0 & \frac{1-\nu}{2} 
%	\end{bmatrix}
%		 \begin{bmatrix} 
%		\epsilon_{xx}  \\
%		\epsilon_{yy}  \\
%		2 \epsilon_{xy} 
%	\end{bmatrix},  \label{stress_strain_relation}
%\end{equation}
%
% \begin{align}
% 	\epsilon_{xx} &=  u_{x,x}  \nonumber \\ 
%	\epsilon_{yy} &=  u_{y,y} \label{strain_displacement_explicit} \\ 
%	\epsilon_{xy} & = \frac{1}{2} (u_{x,y} +  u_{y,x}) \nonumber
%\end{align}
%
%  \begin{align}
% 	\frac{ \partial \left[ \frac{E}{1-\nu^2} \left(\epsilon_{xx} + \nu \epsilon_{yy} \right) \right]} { \partial x} +  \frac{ \partial \left[ \frac{E (1-\nu)}{(1-\nu^2)} \epsilon_{xy} \right]} { \partial y}  &= 0 \\
%	 \frac{ \partial \left[ \frac{E (1-\nu)}{(1-\nu^2)} \epsilon_{xy} \right]} { \partial x} + \frac{ \partial \left[ \frac{E}{1-\nu^2} \left( \nu \epsilon_{xx} +  \epsilon_{yy} \right) \right]} { \partial x} &= 0  \nonumber
%\end{align}